\documentclass[aps,reprint,superscriptaddress,nofootinbib,showpacs,amsmath,amssymb]{revtex4-1}

\usepackage{latexsym,graphicx,slashed,bm,dcolumn}

\newcommand{\bi}{\bibitem}

\def\bef{\begin{figure}}
\def\eef{\end{figure}}

\newcommand{\be}[1]{\begin{equation}\label{#1}}
\newcommand{\beq}{\begin{equation}}
\newcommand{\ee}{\end{equation}}
\newcommand{\beqn}[1]{\begin{eqnarray}\label{#1}}
\newcommand{\eeqn}{\end{eqnarray}}
\newcommand{\bd}{\begin{displaymath}}
\newcommand{\ed}{\end{displaymath}}

\def\lsim{\raise0.3ex\hbox{$\;<$\kern-0.75em\raise-1.1ex
e\hbox{$\sim\;$}}}
\def\gsim{\raise0.3ex\hbox{$\;>$\kern-0.75em\raise-1.1ex
\hbox{$\sim\;$}}}
\def\simlt{\mathrel{\lower2.5pt\vbox{\lineskip=0pt\baselineskip=0pt
           \hbox{$<$}\hbox{$\sim$}}}}
\def\simgt{\mathrel{\lower2.5pt\vbox{\lineskip=0pt\baselineskip=0pt
           \hbox{$>$}\hbox{$\sim$}}}}
\def\unity{{\hbox{1\kern-.8mm l}}}
\newcommand{\ov}{\overline}
\renewcommand{\to}{\rightarrow}

\newcommand{\dm}{\epsilon }   

\newcommand{\nbar}{\tilde{n}}
\newcommand{\tn}{\tilde{n}}

\renewcommand{\to}{\rightarrow}

\def\lsim{\mathrel{\mathop  {\hbox{\lower0.5ex\hbox{$\sim$}
\kern-0.8em\lower-0.7ex\hbox{$<$}}}}}
\def\gsim{\mathrel{\mathop  {\hbox{\lower0.5ex\hbox{$\sim$}
\kern-0.8em\lower-0.7ex\hbox{$>$}}}}}

\def\cB{{\cal B}}

\def\cN{{\cal N}} 
\def\cM{{\cal M}}
\def\cO{{\cal O}}

\def\sq{{\it s}q}
\def\su{{\it s}u}
\def\sd{{\it s}d}
\def\ss{{\it s}s}
\begin{document}

\title{Neutron-antineutron Oscillation and Baryonic Majoron: \\
Low Scale Spontaneous Baryon Violation}

\author{Zurab~Berezhiani}
\affiliation{Dipartimento di Fisica, Universit\`a dell'Aquila, Via Vetoio, 67100 Coppito, L'Aquila, Italy} 
\affiliation{INFN, Laboratori Nazionali Gran Sasso, 67010 Assergi,  L'Aquila, Italy}





\begin{abstract}
We discuss a possibility that baryon number $B$ is spontaneously broken at low scales, 
of the order of MeV or even smaller,   
so that the neutron--antineutron oscillation can be induced at the experimentally accessible 
level.  An associated Goldstone particle -- baryonic majoron, can have observable 
effects in neutron to antineutron transitions in nuclei or dense nuclear matter.  
By extending baryon number  $B-L$ symmetry, baryo-majoron can be identified with 
the ordinary majoron associated with the spontaneous breaking of lepton number, 
with interesting implications for neutrinoless $2\beta$ becay with the majoron emission. 
 We also discuss a hypothesis 
that baryon number can be spontaneously broken by the QCD itself via the six-quark condensates. 
 \end{abstract}

\maketitle


\section{Introduction} 

There is no fundamental principle that can prohibit to neutral particles as are the neutron 
or neutrinos to have a Majorana mass envisaged long time ago by Ettore Majorana 
 \cite{Majorana}. 
 Nowadays the neutron is known to be a composite fermion having a Dirac nature 
 conserving baryon number. As for the neutrinos, theorists prefer to consider them 
 as Majorana particles  though no direct experimental proofs for this were obtained yet 
(e.g. the neutrinoless double-beta decay).  
On the other hand, it is not excluded that the neutron $n$, along the Dirac mass term 
$m\,\ov{n} n$, with $m \approx 940$ MeV, has also a Majorana mass term 
$\dm\, nCn + {\rm h.c.} = \dm\, \ov{n}\,  \tilde{n} +  {\rm h.c.} $, with $\dm \ll m$,   
 which mixes the neutron and antineutron states   
(here $C$ is charge conjugation matrix and $\tn = C\ov{n}^t$ is the antineutron field).  
This mixing induces a very interesting phenomenon of neutron--antineutron oscillation, 
$n \leftrightarrow  \tilde n$  suggested by Kuzmin \cite{Kuzmin:1970nx}. 
First theoretical scheme for $n-\nbar$ oscillation were suggested in Ref. 
\cite{Mohapatra:1980qe}, 
followed by other types of 
models as e.g. \cite{Babu:2001qr,Berezhiani:2005hv,Babu}. 

  Clearly, existence of the Majorana mass of the neutron  would 
violate the conservation of baryon number $B$ by two units
(analogoulsy, Majorana masses for neutrinos violate  lepton number $L$ by two units).   
If $B$  and $L$  were exactly conserved,  the phenomena
like proton decay, $n - \nbar $  oscillation or neutrinoless $2\beta$ decay 
would be impossible.
Experimental limits on matter stability tell that $B$-violating processes must be very slow:  
lower bounds on the lifetime of the nucleons  
(and of stable nuclei) land between $10^{30} - 10^{34}$ yr \cite{PDG}. 
On the other hand, we have a strong theoretical argument that baryon number 
must be indeed violated in some processes -- the existence of matter itself. 
Without $B$-violation  no primordial baryon asymmetry could be generated after 
inflation and so  
the universe would remain baryon symmetric and thus almost empty of matter.   
Primordial baryogenesis in the Early Universe 
maybe related to the same $B$-violating physics that induces neutron-antineutron mixing.  
As was shown by Sakharov, $B$-violating processes which break also CP and which 
were out of equilibrium at some early cosmological epoch, can generate  non-zero 
baryon number in the universe \cite{Sakharov:1967dj}.
(In  modern theoretical scenarios,  
$B-L$ violation is indispensable and also sufficient \cite{Kuzmin:1985mm}.)
It is interesting to note that $n-\nbar$ oscillation implies breaking of P and CP along 
with $B-L$ violation, so that 
two of three Sakharov's conditions for baryogenesis are automatically 
satisfied \cite{Berezhiani:2015uya}. 
Hence, discovery of neutron-antineutron 
oscillation would make it manifest that these underlying physics, 
based e.g. on models \cite{Mohapatra:1980qe,Berezhiani:2005hv,Babu}, 
contain CP violating terms which could be 
at the origin of the baryon asymmetry of the Universe. 

The structure of the Standard model describing the known particles and 
their interactions nicely explains why the $B$ and $L$ violating processes are suppressed.  
Under the standard gauge group $G = SU(3) \times SU(2) \times U(1)$,  
the left-handed quarks and leptons transform as  
iso-doublets $q_L = (u, d)_L$, $l_L = (\nu, e)_L$ while 
the right-handed ones are iso-singlets
$u_R$, $d_R$, $e_R$.
(For simplicity, hereafter we omit the symbols L (left) and R (right) 
as well as the internal gauge, spinor and family indices;  
antiparticles will be termed as $\tilde{q}$, $\tilde{l}$, etc. and 
the charge conjugation matrix $C$ will be omitted.)   
As usual, we assign a global lepton charge $L = 1$ to leptons and 
a baryon charge $B = 1/3$ to quarks, so that baryons composed of three valent quarks 
have a baryon number $B=1$. 

However, $L$ and $B$ are not perfect quantum numbers.
They are related to accidental global symmetries possessed by the 
Standard Model Lagrangian at the level of  renormalizable couplings 
({\sl no renormalizable coupling can be written that could violate them)}. 
However, they  can be explicitly broken by higher dimension 
({\sl non-renormalizable})  operators suppressed by large mass scales  
which may be related to the scales of new physics 
beyond the Standard Model  \cite{Weinberg}.
E.g., grand unified theories (GUTs) 
introduce new interactions that transform quarks into leptons and thus induce 
effective $D=6$ operators $\frac{1}{M^2} qqql$, etc. 
which lead to the proton decays like $p \to \pi e^+$, $p \to K \nu$ etc. 
These decay rates are suppressed by the GUT scale 
$M\geq 10^{15}$~GeV 
which makes them compatible with the existing experimental limits \cite{PDG}.  

The lowest dimension operator, D=5, is related to leptons and it 
violates the lepton number by two units  \cite{Weinberg}: 
\be{ll}
\cO_5 =  \frac{1}{M} l \phi l \phi   \, ~~~~~~ (L=2)
\ee
where $\phi$ is the Higgs doublet. After inserting the Higgs VEV  $\langle \phi\rangle$, 
this operator yields small Majorana masses for neutrinos, 
$m_\nu \sim \langle \phi\rangle^2/M$, 
and induces oscillations between different neutrino flavors. 
Interestingly, the experimental range of the neutrino masses, 
$m_\nu \sim 0.1$~eV or so,  also favors the GUT scale $M\sim 10^{15}$~GeV as 
a natural scale of these operators. 

The neutron - antineutron mass mixing,  $\dm (\ov{n} \,\nbar  + {\rm h.c.})$,  
violates the baryon number  by two units. 
It  can be related to the effective D=9 operators 
involving six quarks which in terms of the Standard model fragments 
 $u=u_R$, $d=d_R$ and $q = (u, d)_L$  read as
 %
 \be{nn}
 \cO_9  = \frac{1}{\cM^5} \big (udd udd \,  + \,  udd qqd \, + \, qqd qqd \big)     ~~~ (B=2)
\ee
 where $\cM$ is some large mass scale.  
 These operators can have different convolutions of the Lorentz, color and weak isospin indices 
 which are not specified.  
 (Needless to say, the combination $qq$ in second term in (\ref{S}) 
must be in a weak isosinglet combination, 
$qq =\frac12 \epsilon^{\alpha\beta} q_{\alpha}q_{\beta} = u_L d_L $ where 
$\alpha,\beta= 1,2$ are the weak $SU(2)$ indices, while in the third term $qq$ can be taken
in a weak isotriplet combination as well.)   
More generally, 
having in mind  that all quark families can be involved, 
these operators give rise 
to mixing phenomena  also  for other neutral baryons, e.g.  
oscillation of the hyperon $\Lambda$ into the antihyperon $\tilde\Lambda$.
 
If the scale $\cM$ is taken of the order of the GUT scale, 
as one takes for the proton decaying  $D=6$  operators $\cO_6$ 
or for $D=5$ neutrino mass operator $\cO_5$ (\ref{ll}), 
the effects of $n-\tilde n$ mixing would become vanishingly small. 
On the other hand, the GUT scale is not really favored by 
the primordial baryogenesis. 
The latter preferably work at smaller scales, in the post-inflation epoch. 
An adequate scale for baryogenesis in the context of $\Delta B = 2$ models 
can be  as small $\cM \sim 1$~PeV \cite{Babu}.


Taking into account that the matrix elements of  operators  $\cO_9$ 
between the neutron states are of the order of
$\Lambda_{\rm QCD}^6 \sim 10^{-4}$ GeV$^6$, modulo the  Clebsch coefficients $O(1)$,   
one can estimate: 
\be{dm}
\dm \sim  \frac{\Lambda_{\rm QCD}^6}{\cM^5} \sim   \left(\frac{1 \, {\rm PeV} }{\cM}\right)^5 \times 
10^{-25} \, {\rm eV}	\; .
\ee
The  coefficients of matrix elements $\langle \nbar \vert \cO_9 \vert n\rangle$ 
for different Lorentz and color structures of operators (\ref{nn}) were studied in ref. \cite{Rao} 
but we do not concentrate here on these particularities and take them as $O(1)$ factors.  
%
In the presence of mixing $\dm (\ov{n} \,\nbar  + {\rm h.c.})$,  
the neutron mass eigenstates become  two Majorana states 
with the masses $m+\dm$ and $m-\dm$, respectively 
$n_+ = \sqrt{1/2} (n + \tilde n)$ and $n_- = \sqrt{1/2} (n - \tilde n)$.  
 The characteristic time of  $n \leftrightarrow  \tilde n$ oscillation 
is related to their mass splitting, $\tau = \dm^{-1}$.

The experimental limit $\tau > 0.86 \times 10^8$~s (90~\%~C.L.) 
obtained by a search of $n-\tilde n$ oscillation with cold  neutrons 
freely propagating in the conditions of suppressed magnetic field \cite{Grenoble} 
implies $\dm < 7.7 \times 10^{-24}$~eV. 
On the other hand, $n-\tn$ mixing inside the nuclei must destabilize the latter \cite{Chetyrkin}. 
In fact, operator (\ref{nn}) induces annihilation processes of two nucleons into pions, 
$N N \to \pi$'s, which transform nucleus with atomic number $A$ into the nucleous 
with $A-2$ with emission of pions with total energy roughly equal to two nucleon masses. 
Interestingly,  nuclear stability limits translated to the free $n-\tn$ oscillation time  
are not far more stringent than direct experimental limit \cite{Grenoble}. 
E.g. the Iron decay limit implies $\tau > 1.3 \times 10^8$~s \cite{Soudan} while  
the Oxygen one $\tau > 2.7 \times 10^8$~s \cite{Super-K}. 
Hence, one can conclude that $n-\tilde n$ oscillation may test the underlying physics 
up to cutoff scales $\cM \sim 1$ PeV, also having in mind possible  
increase of the experimental  senisitivity by an order of magnitude. 
For the discussion of the present status of $n-\tilde n$ oscillation and  
future projects for its search see e.g. refs. \cite{Projects}.   

\medskip 

One can envisage a situation when baryon number is broken not explicitly but spontaneously.  
In particular, one can consider a situation when baryon number associated with an exact global symmetry  $U(1)_B$ 
is spontaneously broken by a complex scalar field  $\chi$ with $B=2$
which breaking also induces the Majorana mass term for the neutron. 
 Clearly,  spontaneous breaking of global $U(1)_B$ gives rise to 
a Goldstone boson $\beta$  
which can be coined as the baryonic majoron, or baryo-majoron, 
in analogy to the majoron associated with the spontaneous breaking of global lepton 
symmetry $U(1)_L$   \cite{Chikashige:1980ui} and widely exploited  in neutrino physics.

In fact, spontaneous baryon violation in the context of $n-\nbar$ oscillation and 
the physics of the baryonic majoron was previously discussed 
in ref. \cite{Barbieri}, in the context of the model \cite{Mohapatra:1980qe}. 
Spontaneous $B$-violation was discussed  also in ref.  \cite{Dvali:2005an}, 
in terms of the operator $qqql$ ($B=1$).   
The associated Goldstone boson was named as bary-axion, 
for respect of the electroweak anomaly of $U(1)_B$. 
 
In this paper we discuss the possibility of spontaneous $B$-violation at very low scales, 
$< 1$~MeV or so, in which case the baryo-Majoron can have observable consequences, 
inducing nuclear decay via the Majoron emission, related to transition $n \to \nbar + \beta$.   
Global baryonic symmetry can be naturally extended to $U(1)_{B-L}$ in which case 
its spontaneous breaking scale must be relevant also for the neutrino Majorana masses, 
and the baryonic and leptonic Majorons become in fact the same particle, just the Majoron. 
In this context, we briefly discuss implications for leptonic sector as e.g. neutrinoless 
$2\beta$ decay with Majoron emission and atsrophysical implications of the Majoron. 
We also shall discuss a rather unusual possibility when baryon number is broken 
by six quark condensates $\langle udd udd \rangle$ and its possible implications.

\section{Seesaw for  { $n-\nbar$}  mixing}


The contact (nonrenormalizable)  $L$ and $B$ violating terms like (\ref{ll}) and (\ref{nn}) 
can be induced in the context of renormalizable theories after decoupling of some heavy 
particles.  In particular, leptonic operator (\ref{ll}) can be induced 
in the context of seesaw  mechanism 
 which involves gauge singlet fermions $N_{(R)}$, 
so called right-handed (RH) neutrinos,  
with large Majorana mass terms $\frac12 M_N N^2 + {\rm h.c.}$ 
explicitly violating $L$. 
Then at energies $E \ll M$, operator (\ref{ll}) emerges from the 
Yukawa couplings $ \phi \ov{N} l  + {\rm h.c.}$ after integrating  out 
the heavy neutrinos  $N$, with $M\sim M_N$. 
Then, modulo Yukawa constants, we have for the Majorana masses of neutrinos  
\be{mnu-f}
m_\nu \sim \frac{v^2}{M_N} \sim  \left(\frac{10^{14}~\rm GeV}{M_N} \right) \times 0.1 ~{\rm eV}
\ee
 
One can also discuss a simple seesaw-like scenario for generation of terms (\ref{nn}),  
along the lines suggested in ref. \cite{Berezhiani:2005hv}. 
%
%
Let us introduce a gauge singlet Weyl fermion (or fermions) $\cN_{(R)}$, 
sort of  ``RH neutron",  
and a color-triplet scalar $S$, with mass $M_S$, having precisely the same gauge 
quantum numbers as the right down-quark $d_{(R)}$. 
Consider the Lagrangian terms 
\be{S}
S \, u \, d  + S \, q \, q  +  S^\dagger  d \, \cN + \frac12 M_{\cN} \cN^2 
 +  {\rm h.c.}   
\ee 
where $qq$ in second term is  in a weak isosinglet combination, 
$qq =\frac12 \epsilon^{\alpha\beta} q_{\alpha}q_{\beta} = u_L d_L $ where 
$\alpha,\beta= 1,2$ are the weak $SU(2)$ indices
(we omit the charge conjugation matrix $C$ and Yukawa constants $\sim 1$). 
One can prescribe $B=-2/3$ to $S$  and $B =-1$ to $\cN$, 
so that the Yukawa couplings in (\ref{S}) respect the baryon number 
which is  explicitly violated by Majorana mass $M_{\cN}$. 
Then, at energies $E \ll M_S,M_{\cN}$, 
operators $\cO_9$ 
of eq. (\ref{nn}) 
are induced via integrating out 
the heavy states $S$ and $\cN$,  
with  $\cM^5 \sim M_S^4 M_{\cN} $ modulo Yukawa constants in (\ref{S}). 
From the model point of view, the scale $\cM =1$ PeV accessible via $n-\tilde n$ oscillation, 
may correspond to a democratic choice when $M_{\cN} \sim M_S \sim 1$~PeV. 
However, it can be obtained in different situations,  
namely (a) light $S$ and heavy $\cN$, e.g.  $M_S \sim 1$ TeV and $M_{\cN} \sim 10^{18}$ GeV, 
or (b) heavy $S$ and light $\cN$, e.g.  $M_S \sim 10^7$ GeV and $M_{\cN} \sim 100 $ GeV.  
Hence, for the neutron-antineutron mixing mass we have 
\be{deltam-f}
\dm \sim \frac{\Lambda_{\rm QCD}^6}{M_S^4 M_{\cN}} \sim 
\left(\frac{10~\rm TeV}{M_S} \right)^4 \! \left(\frac{10^{14}~\rm GeV}{M_{\cN}} \right)
\times 10^{-25}~{\rm eV}
\ee
Let us notice that the ``heavy neutrino"  $N$ and ``heavy neutron"  $\cN$ 
cannot be the same particle. Otherwise its exchange would induce also 
operators like $udd\nu$ with too low cutoff scale which would 
 induce too fast  proton decay. 
 If they are singlets, they can be divided by 
some discrete symmetries.  Alternatively, one can consider 
$N$  as weak isotriplet and $\cN$ as color octet, in which case 
no mixed mass terms may exist between $N$ and $\cN$ states.   
(In the case of color-octet $\cN$ the scalars $S$ can be taken also as color anti-sextets). 
The exchange via color-octet $\cN$ would generate operators
$\cO_9 \propto (udd)_8 (udd)_8$  with $(udd)_8$ in a color octet combination, 
$(udd)_8 \sim ud \lambda^a d $ 
where $\lambda^a$ are the Gell-Mann matrices.  
Via Fierz Transformation, exchanging $d$ states from the left and right brackets in 
such $\cO_9$, 
the matrix element $\langle n \vert \cO_9 \vert \tilde n \rangle$  
will contribute to the $n-\tilde n$ mixing.
In the context of supersymmetry, such operators can be easily obtained 
via $R$-parity 
breaking terms $u_A d_B d_C$ $(B\neq C)$ in the superpotential, where 
$A,B,C$ are the family indices. 
Taking e.g. a superpotential term $ u d s$ involving 
the up, down and strange RH supermultiplets, 
one obtains the couplings analogous  to $ S ud  + S^\dagger d \cN$ of (\ref{S})  
with $S$ being the strange squark and $\cN$ being gluino 
with a Majorana mass $\tilde M$. This is because the gluino 
may have flavor-changing coupling between quark and squark states, 
namely between $d$-quark and $s$-squark. 
Needless to say, in this scheme somewhat bigger mixing mass would be generated 
for hyperons, between $\Lambda$ and $\tilde \Lambda$, via flavor diagonal gluino coupling 
between $s$-quark and $s$-squark.  
However, $\Lambda - \tilde \Lambda$ 
mixing is much more difficult for the experimental detection (though it maybe more efficient 
in the dense nuclear matter in the neutron stars where hyperons 
can emerge as natural occupants). In any case, $\Lambda-\tilde \Lambda$  mixing would also 
induce nuclear instability via two nucleon annihilation processes with Kaon emission, 
$N+N\to K + K$ etc.    

The interesting link between seesaw mechanisms for generation 
of the neutrino and neutron Majorana masses is the following. 
In parallel to usual leptogenesis scenario \cite{Fukugita:1986hr}
due to the heavy neutrino decays $N\to l\phi$  
producing lepton number which then is redistributed to baryon number 
via $B-L$ conserving sphaleron effects \cite{Kuzmin:1985mm}, also baryogenesis 
can take via the heavy `neutron' decays $\cN \to udd$ mediated via color-triplet scalar $S$ 
which can directly produce the baryon number  of the universe. 



\begin{figure}[t]
\begin{center}
\vspace{-1.2cm}
\includegraphics[width=8cm]{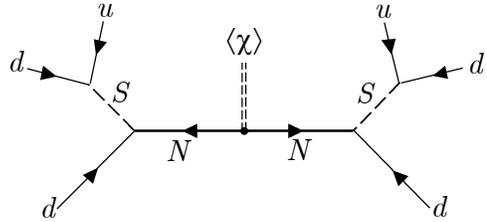}
\vspace{-1.4cm}
\caption{
\label{fig1}
Diagram generating  $n - \tilde n$ mixing 
via exchange of $\cN$ state which gets a large Majorana mass $M_{\cN} \sim \langle \chi\rangle $ 
after $U(1)_B$ symmetry breaking.}
\end{center}
\end{figure}

\medskip 

Let us consider now  a situation when baryon number is broken not explicitly but spontaneously.  
Namely, let us assume that baryon number associated with an exact global symmetry  $U(1)_B$, 
and it is spontaneously broken by a complex scalar field  $\chi$ ($B=2$) 
once the latter gets a VEV $\langle \chi \rangle = V$. 
%
The seesaw Lagrangian (\ref{S}) in this case is modified as 
\be{S-chi}
S \, u \, d  + S \, q \, q  +  S^\dagger  d \, \cN + \chi \cN^2 
 +  {\rm h.c.}   
\ee 
The VEV of $\chi$  induces the Majorana mass 
 to the RH neutron $\cN$  through the Yukawa coupling  $\chi \cN^2 $, 
with $M_{\cN} \sim V$.  
Hence, operator $\cO_9$ emerges after the spontaneous baryon violation 
as shown on Fig. \ref{fig1}, and  $n-\nbar$ mixing parameter $\dm$     
is inversely proportional to the baryon symmetry 
breaking scale $V$.

The scale $V$ can be related also to the breaking of lepton number if one extends 
global symmetry $U(1)_B$ to $U(1)_{B-L}$ and assumes  
that the neutrino Majorana masses emerge from the usual seesaw Lagrangian
\be{seesaw} 
\phi \ov{N}l  + \chi^\dagger N^2 + {\rm h.c.} 
\ee
Since the neutrino masses (\ref{mnu-f}) point twoards $U(1)_{B-L}$ breaking scale 
$V\sim 10^{14}$~GeV, then $n-\nbar$ oscillation, according to  (\ref{deltam-f}),  
can be within the experimental reach if color triplets $S$ have masses 
in the range $M_S \sim 10$~TeV, potentially within the reach for the LHC run II. 

Spontaneous breaking of global $U(1)_B$ or  $U(1)_L$  gives rise to 
a Goldstone boson $\beta$, baryo-majoron or lepto-majoron.  
These two can be the same particle, simply a majoron, once the 
global symmetry is promoted to $U(1)_{B-L}$. However, in practice 
very large scale of symmetry breaking renders such majoron(s) unobservable  
experimentally and without any important astrophysical consequences. 
In the following section we discuss models where the global symmetry 
breaking scale can be rather small, $< 1$~MeV or less, in which case 
the majoron interactions with the neutron and with neutrinos 
could have observable experimental and astrophysical consequences.    

\section{Low scale seesaw model} 

Is it possible to built a consistent model in which baryon number, or $B-L$,    
spontaneously breaks at rather low scales in which case the majoron couplings 
to the neutrinos and to the neutron can be accessible for the laboratory search?  
This can be  obtained by a simple modification of the above considered model. 
 
\begin{figure}[t]
\begin{center}
\vspace{-1.cm}
\includegraphics[width=8cm]{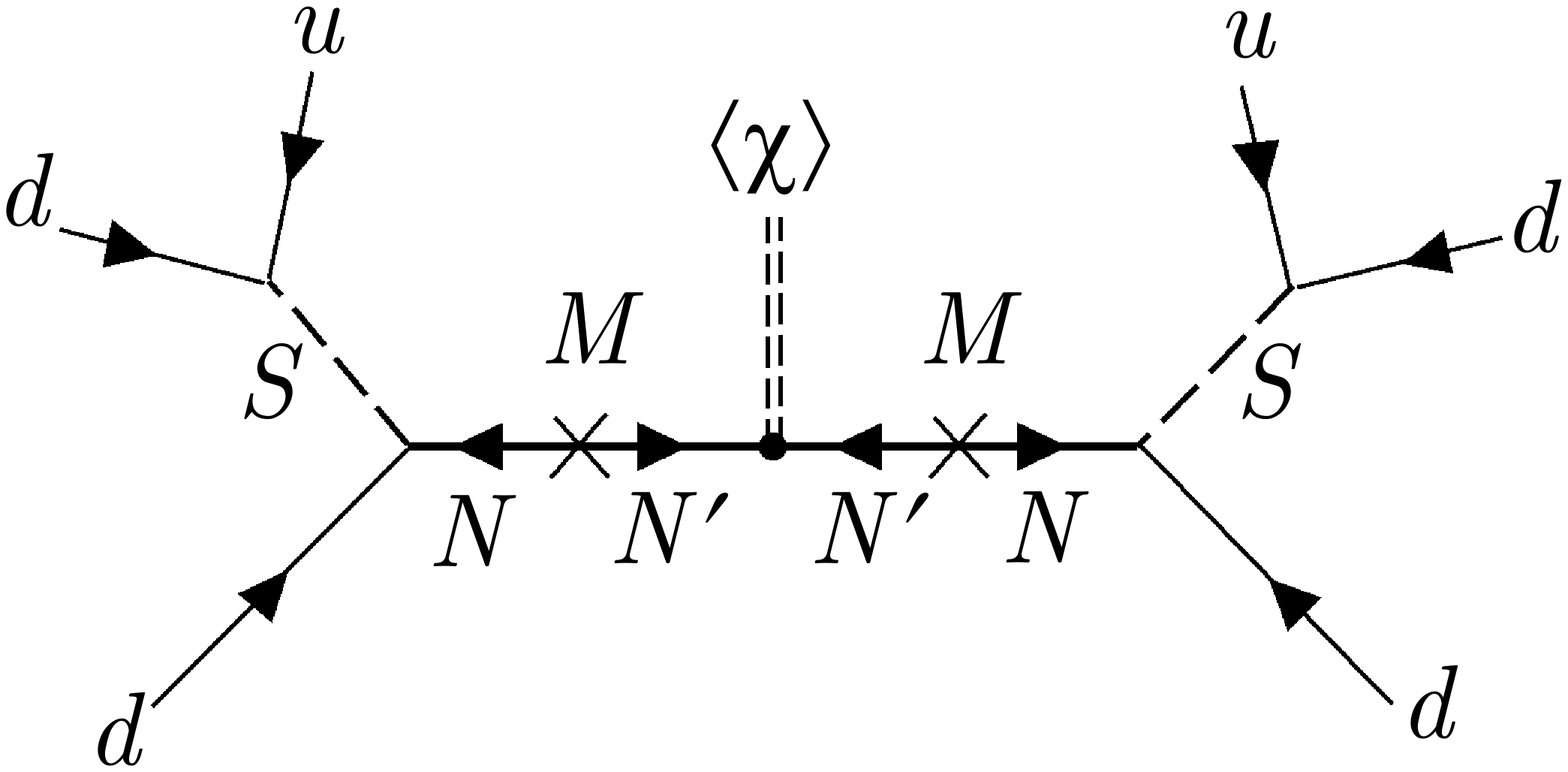} \\
\vspace{-2.4cm}
\includegraphics[angle=270,width=8cm]{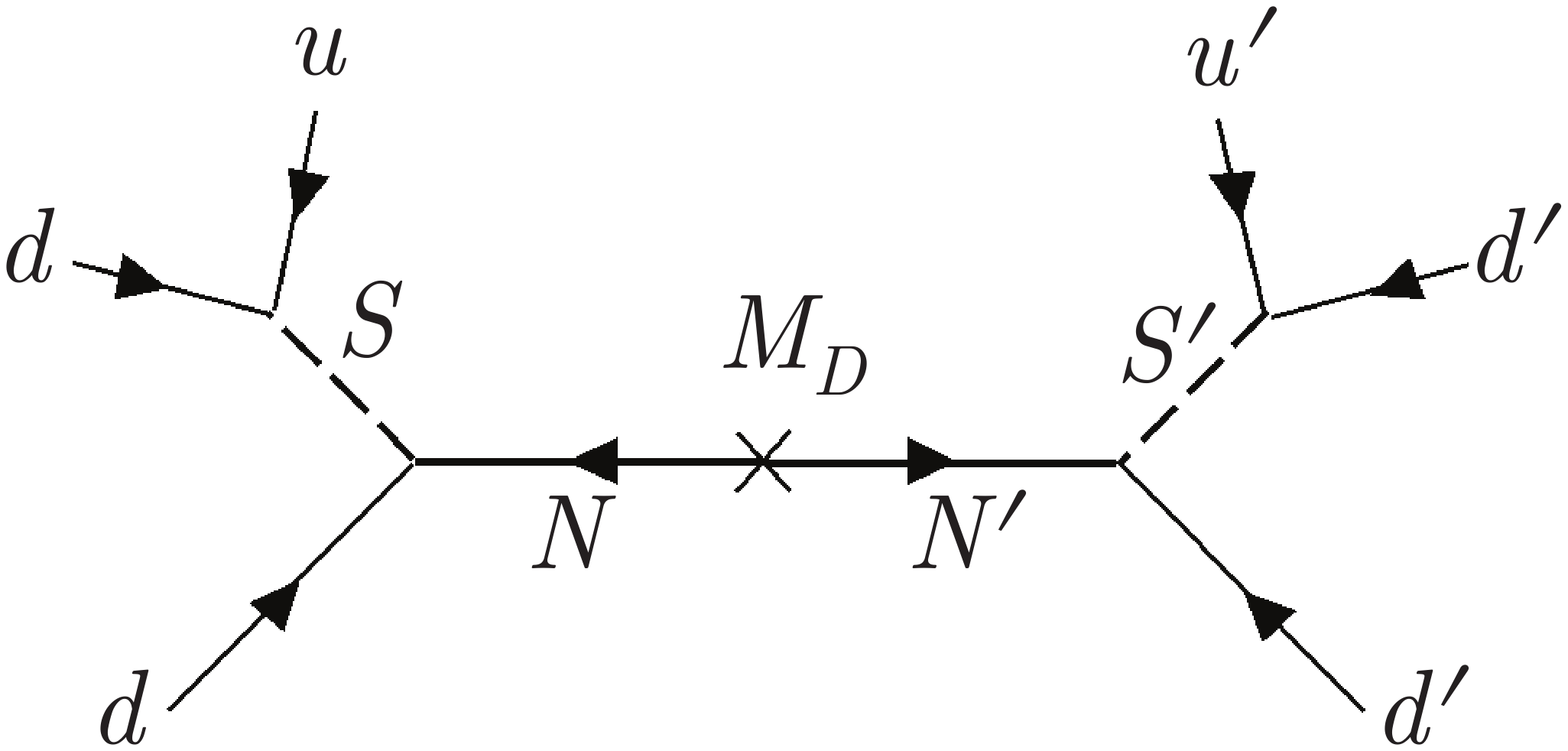}
\vspace{-1.2cm}
\caption{
\label{fig2}
Upper diagram generates  $n - \tilde n$ mixing in low scale baryo-majoron model via exchange of 
heavy Dirac fermion $\cN+\cN'$  when $\cN$ and $\cN'$ get a small Majorana mass 
$\tilde{M}, \tilde{M}' \sim \langle \chi\rangle $. In the presence of  mirror sector containing 
the twin quarks $u',d'$ connected to $\cN'$, lower diagram would generate $n-n'$ mixing 
which conserves the combination of Baryon numbers $B-B'$, without insertion of $\chi$ field. 
}
\end{center}
\end{figure}

Let us introduce along with the Weyl fermion $\cN$ with $B=-1$,  
also another guy $\cN'$ with $B=1$. These two together form a heavy 
Dirac particle with a large mass $M_D$. 
On the other hand, both $\cN$ and $\cN'$ can be coupled to  
scalar $\chi$ ($B=2$) and get the Majorana mass terms 
from the VEV of the latter, $\tilde M, \tilde{M}' \sim \langle \chi \rangle$, 
 which can be much less than the Dirac Mass $M_D$.
The relevant Lagrangian terms now read: 
\be{NNpr}
S u d  +  S q q  +  S^\dagger  d \, \cN + 
M_D \cN \cN' +  \chi \cN^2 +  \chi^\dagger \cN^{\prime 2}  + {\rm h.c.} 
\ee
In this way,  diagram shown in Fig. \ref{fig2},
after integrating out the heavy fermions $\cN + \cN'$, induces  $D=10$ operators 
\be{nn-chi}
\cO_{10}  \sim \frac{\chi^\dagger}
{M_D^2 M_S^4} \big( udd udd \,  + \dots )
 ~~~ (B=0)
\ee
where dots stand for other field combinations present in (\ref{nn}). 
Now, assuming that the field $\chi$ is light, one can consider the matrix element 
directly of $\cO_{10}$ between the $n$ and $\nbar$ states.
Thus, at low energies these operators reduce to the neutron Yukawa couplings with scalar $\chi$, 
\be{Yn}
Y_n \chi^\dagger n^T C n + {\rm h.c.}, 
\ee
with the coupling constant
\be{Yuk}
Y_n \sim \frac{\Lambda_{\rm QCD}^6}{M_D^2 M_S^4} \sim 
\left(\frac{100~\rm TeV}{M_D} \right)^2 \left(\frac{10~\rm TeV}{M_S} \right)^4 
\times 10^{-30}
\ee
Thus, once the baryon number is broken by the VEV $\langle \chi \rangle $, 
%
$n-\tilde n$ mixing emerges with 
$\dm_{n\nbar} = Y_n \langle \chi \rangle \sim \Lambda_{\rm QCD}^6 \tilde{M}'/(M_S^4M_D^2)$, 
or 
\be{n-nbar-mix}
\dm_{n\nbar} \sim 
\left(\frac{100~\rm TeV}{M_D} \right)^2 \left(\frac{10~\rm TeV}{M_S} \right)^4 
\left(\frac{\tilde{M}'}{1~\rm MeV}\right) \times 10^{-24}~ {\rm eV} .
\ee
%
Taking e.g. $M_S \sim 10$~TeV and $M_D \sim 100$~TeV, 
then $\dm_{n\nbar} \sim 10^{-24}$~eV would require 
$\langle \chi \rangle \sim 1$ MeV or so.\footnote{The following remark is in order. 
In general, operators  (\ref{nn-chi}) can contain parts which respect and which 
do not respect P-parity. 
Therefore, taking matrix element $\langle n \vert \cO_{10} \vert \nbar \rangle$, 
in addition the coupling (\ref{Yn}) 
one can have also P-invariant coupling $Y'_n \chi^\dagger n^T C\gamma^5 n + {\rm h.c.}$.
Only the former term violating P will be relevant for $n-\nbar$ oscillation after 
non-zero $\langle chi \rangle$ breaks $B$   
\cite{Berezhiani:2015uya}. For the majoron interactions also the latter term  
would be relevant which now can have both P-invariant and P-violating couplings. 
For simplicity we shall not discuss it in the following.}

Low scale baryon number violation was suggested in Ref. \cite{Berezhiani:2005hv}, 
in a model which was mainly designed for inducing  
neutron -- mirror neutron oscillation $n-n'$. 
 This model treats $\cN$ and  $\cN'$ states  symmetrically: their Majorana masses 
 $\tilde{M}$ and $\tilde{M}'$ are equal, while in addition to couplings (\ref{NNpr}), 
 there are terms that couple $\cN'$ to  $u', d'$ and $S'$ states
 from hidden mirror sector  with a particle content identical to that of ordinary one 
 (for review, see e.g. \cite{Berezhiani:2003xm}).  
Namely, the lower diagramm of Fig. \ref{fig2}. induces $n-n'$ mixing  with 
\be{nn'}
\dm_{nn'} \sim \frac{\Lambda_{\rm QCD}^6}{M_D M_S^4} 
\sim \left(\frac{100~\rm TeV}{M_D} \right) \left(\frac{10~\rm TeV}{M_S} \right)^4 
\times 10^{-16}~{\rm eV}
\ee
which corresponds to $n-n'$ oscillation time $\tau_{nn'} \sim 10$~s. 
Hence, in this case  $n-n'$ mixing should be a dominant effect, conserving a combined 
baryon number $\bar B = B -B'$ between ordinary and mirror sectors, 
while $n-\nbar$ mixing which breaks $\bar B$ is suppressed by the ratio $\tilde{M}/M_D$:  
\be{delta-ratio}
\dm_{n\nbar}  = \frac{\tilde{M}}{M_D} \dm_{nn'}
\ee
As a matter of fact, $n-n'$ mixing can indeed be much larger than $n-\nbar$.  
Existing experimental limits on $n-n'$ transition \cite{nnpr-exp} 
allow the neutron$-$mirror neutronoscillation time to be less than the neutron lifetime, 
with interesting implications for astrophysics and particle phenomenology 
\cite{Berezhiani:2005hv,nnpr}.

 \medskip

Let us discuss now the couplings of the majoron $\beta$ 
which is a Golsdtone component 
of the $\chi$ scalar, $\chi =  \frac{1}{\sqrt2 }(V + \rho)\exp(i\beta/V) $, 
where $\rho$ denotes a massive (Higgs) mode of $\chi$ with a mass $\sim V$.
(We assume here that the VEV $\langle \chi \rangle = V/\sqrt2$ emerges via 
negative mass$^2$ term in the potential of $\chi$.)  
Both $\rho$ and $\beta$ are coupled non-diagonally  between the $n$ and $\tilde n$ states, 
$g_{\beta nn} \ov{n} (\rho + i \beta \gamma_5) \tilde n  + {\rm h.c.} $, with 
$g_{\beta nn} = \frac{1}{\sqrt2}Y_n = \dm/V$. 
Observe that the Higgs $\rho$ is coupled to pseudoscalar combination $\ov{n} \nbar$ 
while the  majoron $\beta$ couples to scalar combination $\ov{n} \gamma_5 \nbar$.  
This is related to the fact that the Majorana mass term $\dm \ov{n} \nbar + {\rm h.c.}$ 
breaks $P$ and $CP$ invariances \cite{Berezhiani:2015uya}.

In vacuum the transition $n \to\nbar + \beta $ is suppressed since $n$ and $\tilde n$ have 
equal masses. (We neglect a tiny mass splitting $\dm < 10^{-24}$~eV between two 
Majorana states $n_+$ and $n_-$.)
However, in the nuclei the neutron and antineutron have different effective potentials 
and thus $n \to \tilde n + \beta $ transition becomes possible which  
clearly would lead to the nuclear instability. 
The produced antineutron then annihilates with other spectator nucleons producing 
pions, thus causing the transtition of a nuclei with atomic number $A$ into a nuclei with 
$A-2$ and pions with invariant mass which in principle should be less  
than a mass difference $M_A - M_{A-2}$ between the initial and doughter nuclei 
as far as part of the energy will be taken by the majoron. 
The  decay width  
can be estimated as  $\Gamma  = (g_n^2/8\pi) \Delta E$, where $\Delta E$ 
is a typical energy budget for this transition which depends on nucleus and which 
is typically order 10~MeV. 
Taking into account the the existing experimental limits on the nuclear decay 
$\Gamma^{-1} > 10^{32}$~yr, we get a rough bound $g_n < 10^{-30}$ or so.  
Needless to say, the scalar component $\rho$ with mass order MeV is 
also relevant for the nuclear transitions $n \to \tilde n + \rho $. 

On the other hand,  taking  $\dm < 10^{-24}$~eV, from (\ref{majoron}) we see that 
 $g_n = \dm/V \sim 10^{-30}$  requires $V < 1$~MeV or so. 
$\Gamma^{-1}  \sim  (V/1~{\rm MeV})^2 (10^{-24}~{\rm eV}/\dm)^2 \times 10^{34}$~yr.  
Taking into account that $\dm < 10^{-24}$~eV, 
this exceeds many orders of magnitude the existing experimental limits $\sim 10^{32}$~yr 
\cite{PDG} unless $V < 1$~MeV or so. 


As for the baryo-majoron coupling constant $g_n=\dm/V$, now it can be large enough  
for making $n\to \nbar + \beta$ decay accessible in the experimental search for 
the nuclear destabilisation.  E.g. the nuclear decays at the level 
$\Gamma^{-1}\sim 10^{32}$~yr  can be obtained via $n-\nbar$ oscillation with 
$\delta \sim 10^{-24}$~eV, or via $n\to \nbar + \beta$ decay 
with $g_n = \dm/V \sim 10^{-30}$. Therefore, if $V < 100$~keV the 
former  mechanism becomes suppressed with respect to the latter 
which becomes dominant from the perspectives of the experimental search. 

\begin{figure}[t]
\begin{center}
\includegraphics[width=8cm]{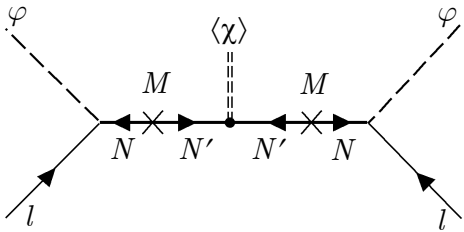}
\vspace{-1.2cm}
\caption{
\label{fig3}
Diagram generating the neutrino majorana masses in a low scale majoron model. }
\end{center}
\end{figure}

Let us remark that in the context of low scale model, with $f_B \leq 1$ MeV, 
baryo-majoron could be the same particle as the usual (leptonic) majoron, 
if one promotes the $U(1)_B$ symmetry to $U(1)_{B-L}$, which is free of anomalies.   
Then the Majorana masses of the neutrinos can be induced, along the lines of the model suggested 
in ref. \cite{Berezhiani:1992cd}, from  the diagram shown in Fig. \ref{fig3}
involving  the following Lagrangian terms 
\be{lepton} 
\phi \, \ov{l} N + M_D NN' +  \chi N^2 +  \chi^\dagger  N^{\prime 2} 
\ee 
where  $N,N'$ are the fermion couples, analogous to $\cN,\cN'$, 
with properly assigned lepton number (or better $B-L$)  
and they have large Dirac masses $M_D$.
Then after integrating out of the heavy states, one obtains an operator 
\be{o6} 
\cO_6 \sim \frac{\chi}{M_D^2} l\phi l \phi
\ee
which at lower energies result in the neutrino Yukawa couplings with the light $\chi$ scalar, 
$Y_\nu \chi \nu^T C \nu + {\rm h.c.} $, where 
\be{Ynu} 
Y_\nu \sim \frac{v^2}{M_D^2} \sim \left(\frac{100~\rm TeV}{M_D} \right)^2 \times 10^{-6}
\ee
Then the neutrino Majorana masses are induced 
with  $m_\nu = Y_\nu \langle \chi\rangle$, or 
%
 \be{nu-mix}
m_{\nu} \sim \frac{v^2}{M_D^2} \langle \chi\rangle \sim 
\left(\frac{100~\rm TeV}{M_D} \right)^2 
\left(\frac{\langle \chi\rangle}{1~\rm MeV}\right) 
\times 1~ {\rm eV} 
\ee

which, taking into account  also uncertainties in the Yukawa constants in (\ref{lepton}),  
naturally fall in the experimental mass range of neutrinos when $M_D \sim 100$ TeV and  
$\langle \chi\rangle < 1$~MeV.\footnote{
Once again, the ``heavy neutrinos"  $N,N'$ and ``heavy neutrons"  $\cN,\cN'$ 
cannot be the same, since in this case their exchange would induce the 
 operators like $udd\nu$ with too low cutoff scale which would 
 lead to dramatically fast  proton decay. }

In this situation, the majoron $\beta$ 
has large enough Yukawa couplings with the neutrinos \cite{Berezhiani:1992cd}, 
with coupling constants $g_{\beta\nu\nu} = m_\nu/V$. Hence, for $V < 1$~MeV 
the  majoron couplings to neutrinos can be rather large, $g_\nu > 10^{-7}$ or so,  
which could be of interest for searching 
the neutrinoless 2-beta decay with the majoron emission \cite{Georgi:1981pg}. 
The present experimental bound on the majoron coupling to $\nu_e$ reads 
$g_{\nu ee} < (0.8 -1.6) \times 10^{-5}$ \cite{Gando:2012pj}. 
In addition, they can bring to interesting effects with interesting applications 
for astrophysics and cosmology as e.g. matter induced neutrino decay  
 or matter induced decay of the majoron itself \cite{Berezhiani:1987gf}, 
blocking of active--sterile oscillations in the early universe 
by the majoron field \cite{Bento:2001xi}, etc. 
Detailed analysis of the astrophysical limits on the neutrino-majoron couplings 
can be found in \cite{Raffelt}. 

The Majoron coupling constant between the neutron and antineutron is 
$g_{n\nbar} = \dm_{n\nbar}/V$. Interestingly, the nuclear decays with majoron 
emission become dominant over majoronless nuclear decays when $V \sim 1$ MeV 
or smaller. 
The parallel of such nuclear decays 
with the neutrinoless 2-beta decays with the majoron emission which also can be 
observable if $V \leq 1$ MeV is interesting.  

One can question the naturality issues  
when having such a small VEVs, $V\sim 1$ MeV, with respect to the electroweak scale 
$M_Z \sim 100$ GeV. If scalar $\chi$ gets a VEV from minimization of its Higgs potential 
with negative mass square, which mass should also be order MeV which gives 
rise a hierarchy problem, why $V \ll M_Z$. This question can be solved if the scale $V$ 
is related to some compositeness scale, e.g. if the scalar $\chi$, even being heavy, with mass 
say $M_\chi \sim 100$ GeV, has the Yukawa couplings with quark-like states,  $\chi \bar Q Q$ 
of some hidden sector  with a confinement scale order MeV.  Then this condensate would 
induce the non-zero VEV to scalar $\chi$, 
$\langle \chi\rangle \sim \langle \ov{Q} Q \rangle/M^2_{\chi}$, and thus the 
Majorana masses for the neutrinos and neutron.

\section{Discussion and outlook} 

 At this point, I am tempted to discuss a less orthodox idea,   
 suggesting that the baryon number could be violated
 by the Standard Model itself, namely  by the strong dynamics of the QCD sector. 
 The conjecture is that along with the basic quark and gluon condensates, 
  $\langle \ov{q}q\rangle$ and $\langle G^2 \rangle$, or higher order operators  
  $\langle \ov{q}\sigma G q\rangle$, $\langle \ov{q}q\ov{q}q \rangle$, 
  there may exist also a  fuzzy six-quark condensates $\langle udd udd \rangle $.  
  These condensates  can be built upon different
  combinations  of left and right $u$, $d$  and perhaps $s$ quarks, and
  may have different convolutions of  the  Lorenz and color indices.  
  One could envisage that they might emerge via 
  attractive forces between the quark trilinears  in color octet combinations.

One interesting possibility can emerge considering that QCD itself could break baryon 
number by two units, by forming a six-quark condensate $\langle udd udd \rangle = \lambda_B^9$. 
Clearly, for experimental compatibility, this condensate must be very fuzzy,  
with a mass parameter $\lambda_B$ order 1 MeV or less. This again would create a hierarchy 
problem, since any condensate in QCD, if it appears, must  have a mass scale order QCD 
scale $\Lambda_{\rm QCD}\sim 200$~MeV. Thus, a fine tuning is required 
of about twenty orders of magnitude. 

Formally,  Vafa Witten theorem \cite{Vafa:1983tf} 
excludes the possibility of baryon number violating condensates in QCD. 
However, this theorem is based in some assumptions which leave some loophole. 
Namely, 
if quarks have masses (as we know our light quarks $u,d,s$ have masses order few MeV), 
the prove is formally valid if the vacuum angle $\Theta$ is exactly zero. 
However,  the vacuum angle might be non-zero: the experimental limit on the electric dipole moment 
of the neutron leads  only  to a theoretical bound $\Theta < 10^{-10}$ or so. 
Then one could envisage that in the possible (but not our) 
world in which $\Theta \sim 1$, the baryon-violating condensates 
could be formed with $V \sim 100$ MeV, however the continuity hypothesis then may 
imply that in the real world the condensate is suppressed by a factor $\Theta^2  < 10^{-20}$ 
which can also explain the smallness of the spontaneous breaking scale $V$.\footnote{ 
For three light flavors $u,d,s$ 
the condensate could appear in flavor singlet combination $\langle uds uds \rangle$, 
and it would induce  the Majorana mass term for the hyperon, i.e. mass mixing 
between the  hyperon and antihyperon states,  
$\dm_\Lambda \ov{\Lambda} \tilde \Lambda + {\rm h.c.}$ } 
 
  Assuming {\it ad hoc} that the six-quark operator   $udd udd $ 
  may have  non-zero VEV in the QCD vacuum, 
  $\langle udd udd \rangle = \cB $, 
  then a Goldstone boson $\beta$ should emerge, the baryo-majoron, 
  as a phase of this condensate, 
  $\cB = \lambda_B^9 \exp(i\beta/f_B)$ where $f_B$ is a respective decay constant.
   However now baryo-majoron becomes a composite field, 
  exactly like pions which are 
  the Goldstone modes of the quark condensate  $\langle \ov{q}q\rangle$ that breaks 
  the chiral $SU(2)_L \times SU(2)_R$ symmetry,  
  $\langle \ov{q}q\rangle = \Sigma \exp(i \tau_a\pi_a/f_\pi)$ with the typical value 
  $\Sigma \simeq (200~{\rm MeV})^3$ and  $f_\pi$ being the pion decay 
  constant. 
   Then one can roughly estimate  the mixing mass between $n - \tilde n$ 
  as $\dm \sim \cB/ (1~\rm GeV)^8$, 
by simply taking scales of the neutron mass and residue and all 
relevant  momenta order 1 GeV and neglecting all combinatorial numerical factors. 
Therefore, if this six-quark condensate is characterized by a mass scale of the order  
of current quark masses, say $\lambda_B \sim 0.3~{\rm MeV}$, 
then we get $\dm  \sim 10^{-23}$~eV, which would correspond 
to $n-\nbar$ oscillation time $\tau_{n\nbar} \sim 10^{8}$~s. 
As for the baryo-majoron, its non-diagonal coupling between $n$ and $\tn$ states 
is related to the value of $\dm$ via Goldberger-Treimann like relation $g_{\beta n} = \dm/f_B$. 
Therefore, for $f_B > 1$ MeV or so, nuclear stability limits versus the neutron decay with 
the majoron emission, $n\to \tn + \beta$ decay,  will be safely respected.

\begin{figure}[t]
\begin{center}
\includegraphics[width=8cm]{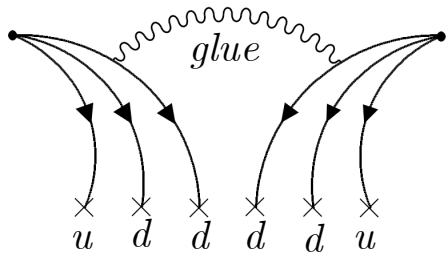}
\vspace{-1cm}
\caption{
\label{fig4}
Diagram generating  the $n - \tilde n$ mixing 
via  baryon-violating six-quark condensate $\langle udd udd \rangle $.}
\end{center}
\end{figure}

An interesting feature of the dynamical baryon violation by the QCD  can be that 
the order parameter $\lambda_B$ could be different in vacuum and in dense nuclear matter, 
i.e. in nuclei or in the interiors of neutron stars. In particular, in dense nuclear matter 
spontaneous baryon violating could occur even if it does not take place in vacuum. 
Or right the opposite, dense nuclear matter could suppress the baryon violating condensates. 
In this case, the search of neutron antineutron 
oscillation with free neutrons and nuclear decay due to neutron antineutron transition 
become separate issues. Namely, it might be possible that the baryon violating condensates 
evaporate at nuclear densities and do not lead to nuclear instabilities while 
for free neutrons propagating in the vacuum they can be at work.    

\bigskip 


\vspace{6mm} 
\noindent {\bf Acknowledgements} 
\vspace{2mm} 

The idea of this work emerged as a result of numerous discussions with  Yuri Kamyshkov. 
I thank Yuri for motivating me to write it down and for a help in preparation of the manuscript. 
I would like to thank Gia Dvali,  Oleg Kancheli, Arkady Vainshtein 
and  Andrea  Addazi  for many valuable discussions.  
The work was reported at Int. Workshop {\it ``NNbar at ESS"}, CERN, 12-13 June 2014.  
I am grateful  Galileo Galilei Institute for Theoretical Physics for the hospitality and partial support 
during the Workshop  {\it ``The Structure and Signals of Neutron Stars, from Birth to Death"}, 
Florence, March 2014,  where the main part of the work was done. 
The work was supported in part by the MIUR 
triennal grant for the Research Projects of National Interest  PRIN 
No. 2012CPPYP7 ``Astroparticle Physics", and in part by 
Rustaveli National Science Foundation grant No. DI/8/6-100/12. 


\end{document}